\documentclass[11pt]{article}

\usepackage[final]{acl}

\usepackage{times}
\usepackage{latexsym}
\usepackage{amsmath}
\usepackage{booktabs}
\usepackage{multirow}
\usepackage{enumerate}
\usepackage{stfloats}
\usepackage{array}
\usepackage{tabularx}
\usepackage[ruled,vlined]{algorithm2e}
\usepackage{listings}
\usepackage[most]{tcolorbox}
\usepackage{xcolor}
\usepackage[T1]{fontenc}
\usepackage{hyperref}
\usepackage{xcolor}

\definecolor{color1}{RGB}{233,230,240} 
\definecolor{color2}{RGB}{194,208,235} 
\definecolor{color3}{RGB}{202,223,184} 
\definecolor{color4}{RGB}{252,230,213} 
\definecolor{colorgray}{RGB}{220,220,220} 
\newtcolorbox{PromptBox}[2]{
    enhanced,
    breakable,              
    pad at break=0pt,
    colback=white,
    colframe=#1!70!black, 
    arc=3pt,
    boxrule=1.5pt,
    left=10pt, right=10pt, top=15pt, bottom=10pt,
    fonttitle=\bfseries\sffamily,
    coltitle=black,
    attach boxed title to top left={xshift=10pt, yshift=-8pt},
    boxed title style={
        colback=#1,
        arc=2pt,
        boxrule=1.5pt,
        left=5pt, right=5pt
    },
    title={#2}
}

\definecolor{myblue}{RGB}{210, 225, 245}
\colorlet{deepblue}{myblue!60!black}

\newtcolorbox{EntityStatBox}[1]{
    enhanced,
    colback=myblue,            
    colframe=deepblue,         
    arc=2pt,                   
    boxrule=1.2pt,             
    fontupper=\small\ttfamily, 
    left=6pt, right=6pt, top=10pt, bottom=6pt,
    fonttitle=\small\sffamily\bfseries,
    coltitle=black,
    attach boxed title to top left={xshift=12pt, yshift=-10pt},
    boxed title style={
        colback=myblue,      
        colframe=deepblue,
        arc=2pt,
        outer arc=2pt,
        boxrule=1.2pt
    },
    title={#1} 
}

\newcommand{\customhrule}{
    \hrule\@width 0.9\textwidth\@height 1pt 
}
\usepackage[utf8]{inputenc}

\usepackage{microtype}

\usepackage{inconsolata}

\usepackage{graphicx}
\hypersetup{pagebackref=true, breaklinks=true, colorlinks, bookmarks=false}

\title{EA-Agent: A Structured Multi-Step Reasoning Agent for Entity Alignment}

\author{
  \textbf{Yixuan Nan}$^{1,2}$, 
  \textbf{Xixun Lin}$^{1,2,}$\thanks{~~Corresponding author.}, 
  \textbf{Yanmin Shang}$^{1,2}$, 
  \textbf{Ge Zhang}$^{3}$, \\ 
  \textbf{Zheng Fang}$^{4}$, 
  \textbf{Fang Fang}$^{1,2}$, 
  \textbf{Yanan Cao}$^{1,2,}$\makeatletter\footnotemark[1]\makeatother \\
  $^1$Institute of Information Engineering, Chinese Academy of Sciences, Beijing, China \\
  $^2$School of Cyber Security, University of Chinese Academy of Sciences, Beijing, China \\
  $^3$School of Information and Intelligent Science, Donghua University, Shanghai, China \\
  $^4$JD.COM, Beijing, China \\
  \texttt{\{linxixun, caoyanan\}@iie.ac.cn}
}

\newtheorem{definition}{Definition}

\begin{document}
\maketitle
\begin{abstract}
Entity alignment (EA) aims to identify entities across different knowledge graphs (KGs) that refer to the same real-world object and plays a critical role in knowledge fusion and integration. Traditional EA methods mainly rely on knowledge representation learning, but their performance is often limited under noisy or sparsely supervised scenarios. Recently, large language models (LLMs) have been introduced to EA and achieved notable improvements by leveraging rich semantic knowledge. However, existing LLM-based EA approaches typically treat LLMs as black-box decision makers, resulting in limited interpretability, and the direct use of large-scale triples substantially increases inference cost. To address these challenges, we propose \textbf{EA-Agent}, a reasoning-driven agent for EA. EA-Agent formulates EA as a structured reasoning process with multi-step planning and execution, enabling interpretable alignment decisions. Within this process, it introduces attribute and relation triple selectors to filter redundant triples before feeding them into the LLM, effectively addressing efficiency challenges. Experimental results on three benchmark datasets demonstrate that EA-Agent consistently outperforms existing EA methods and achieves state-of-the-art performance. The source code is available at \url{https://github.com/YXNan0110/EA-Agent}.
\end{abstract}

\section{Introduction}
Knowledge graphs (KGs) represent real-world knowledge in a structured form and have been widely applied to various knowledge-driven tasks, including question answering~\cite{saxena2022sequence, wang2024knowledge, xu2024retrieval}, personalized recommendation~\cite{jiang2024diffkg, cui2024rakcr, wang2025knowledge}, and knowledge reasoning~\cite{wang2024large, jiang2025kg, liu2026pathmind}. However, due to differences in data sources and construction methodologies, KGs are inherently heterogeneous. 
\par
To enable knowledge fusion across different KGs, entity alignment (EA) plays a crucial role~\cite{zhang2022benchmark}. EA aims to identify equivalent entities in different KGs that refer to the same real-world object, constituting one of the most fundamental techniques for knowledge fusion and a key step toward integrating heterogeneous information~\cite{zeng2021comprehensive, zhu2024survey}. Previous EA methods are mainly based on knowledge representation learning, where entities are aligned by learning embeddings from different KGs and measuring their similarities~\cite{bordes2013translating, wang2018cross}. However, such methods heavily depend on high-quality structural information and sufficient labeled training data, and thus often suffer from limited robustness in scenarios with noise or sparse supervision.
\par
Recently, the emergence of large language models (LLMs) has brought revolutionary advances to a wide range of KG-related tasks~\cite{sun2023think, zhang2024extract, liu2025enhancing}. Owing to their strong semantic understanding and reasoning capabilities, LLMs substantially enhance the modeling of entities and their semantic and relational information~\cite{lin2025llm}. Recently, several studies have incorporated LLMs into EA and achieved state-of-the-art (SOTA) performance on multiple benchmark datasets~\cite{jiang2024unlocking, yang2024advancing, yang2024two}. These LLM-based EA methods typically treat LLMs as the final decision maker, selecting the optimal aligned entity from candidate sets through multi-round voting or reflection mechanisms.
\par
Despite the remarkable improvements achieved by LLM-based EA methods, two key issues remain. \textbf{First, LLM-based EA methods lack interpretability.} Most existing approaches directly feed entities and their triples into LLMs and obtain the alignment results. Consequently, the entire alignment process remains a black-box decision procedure, making it difficult to identify which information is critical to alignment decisions. This opacity hinders error analysis and prevents systematic correction of erroneous alignments. \textbf{Second, the scale of triples poses severe efficiency issues.} Current methods usually input a large number of attribute and relation triples into LLMs, which significantly increases prompt length and inference cost, leading to substantial token consumption. Moreover, many of these triples are redundant or even noisy, which distorts the model’s judgment and further undermines the stability and reliability of the alignment results.

\begin{figure}[t]
  \includegraphics[width=\columnwidth]{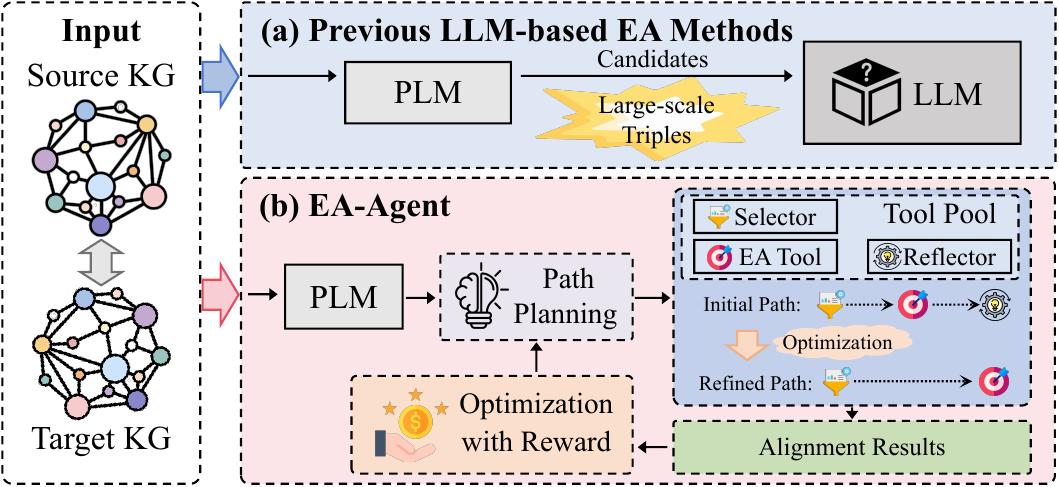}
  \caption{A simple comparison of previous LLM-based EA methods and our proposed EA-Agent.}
  \label{fig:comparison}
\end{figure}

To address the above limitations, we propose \textbf{EA-Agent}, a reasoning-driven agent for entity alignment. As shown in Figure~\ref{fig:comparison}, EA-Agent introduces a new learning paradigm that decomposes EA into a structured reasoning process with multi-step planning and execution. Specifically, we construct a tool pool consisting of triple selectors, an EA tool, and an (optional) reflector. These components together form a structured reasoning process of “\emph{information selection} → \emph{alignment decision} → \emph{result verification}”, enabling the agent to autonomously plan and execute tool invocation paths to produce interpretable and controllable alignment decisions. An \textit{agent optimization mechanism} is further proposed to jointly refine its path planning and alignment behaviors within this structured reasoning process. Moreover, to mitigate the efficiency issues caused by large-scale triples, EA-Agent incorporates \textit{attribute and relation triple selectors} into the tool pool, which filter redundant triples before LLM invocations, significantly reducing token consumption while preserving highly discriminative information. Experimental results on EA benchmarks demonstrate the effectiveness of EA-Agent.

In summary, our contributions are as follows:
\begin{enumerate}[1)]
\item We propose EA-Agent, a novel reasoning-driven agent for EA, which introduces a new learning paradigm that formulates EA as a structured reasoning process with multi-step planning and execution.
\item We design an agent optimization mechanism based on reward-guided offline policy optimization, which enables the agent to iteratively refine its tool planning and alignment behaviors through path rewriting.
\item Extensive experiments on three public benchmark datasets demonstrate that EA-Agent outperforms existing SOTA methods, achieving up to 3.17\% improvement in Hits@1 and consistent gains in MRR.
\end{enumerate}

\section{Related Work}
Existing EA methods are predominantly built upon knowledge representation learning, where entities from two KGs are encoded into a shared embedding space and aligned via similarity computation. These approaches can be broadly categorized into three types: translation-based methods, GNN–based methods, and pretrained language model (PLM)–based methods. 

\textbf{Translation-based EA methods}, such as TransE~\cite{bordes2013translating} and its variants~\cite{lin2015modeling, chen2016multilingual, zhu2017iterative, sun2017cross, sun2018bootstrapping,lin2019guiding}, learn low-dimensional embeddings for entities and relations in different KGs and map them into a unified vector space, where EA is performed via distance minimization. \textbf{GNN-based EA approaches}~\cite{wang2018cross, wu2019relation, li2019semi, wu2019jointly} generate entity embeddings by aggregating neighborhood information. GCN-Align~\cite{wang2018cross} is the first to apply GCN to EA, encoding graph structures and mapping entities into a unified embedding space. KECG~\cite{li2019semi} leverages graph attention networks to down-weight irrelevant neighbors, mitigating noise introduced by heterogeneous KG structures.

With the rise of pretrained language models (PLMs), \textbf{PLM–based EA methods} have attracted increasing attention. These methods leverage the rich semantic knowledge contained in PLMs to encode entities. BERT-INT~\cite{tang2020bert} encodes entity descriptions and interaction information using BERT. TEA~\cite{zhao2023alignment} transforms both attribute and relation triples into unified text sequences and formulates EA as a bidirectional textual entailment task between entities across KGs. More recently, with the emergence of LLMs featuring stronger reasoning abilities, a new line of LLM-based EA approaches has been proposed. ChatEA~\cite{jiang2024unlocking} reformulates KG structures into code-like formats to enhance the understanding of LLMs, and adopts a two-stage alignment strategy to improve accuracy. Seg-Align~\cite{yang2024advancing} introduces sample segmentation to reduce token consumption and enables LLMs to handle large-scale EA. LLMEA~\cite{yang2024two} feeds entity embedding similarities and edit distances into an LLM for reasoning. LLM-Align~\cite{chen2024llm} employs heuristic-based triple importance estimation and a multi-round voting mechanism to stabilize LLM predictions.

Unlike previous work, we propose a reasoning-driven agent that addresses EA through explicit multi-step tool planning and execution. This paradigm facilitates structured and interpretable reasoning, while addressing efficiency issues through triple selection.

\section{Problem Definition}
In this section, we first introduce the formal definitions of knowledge graphs and entity alignment. Based on these definitions, we present our task formulation.
\begin{definition}
(Knowledge Graph) A knowledge graph is defined as $\mathcal{KG}=\{\mathcal{E}, \mathcal{R}, \mathcal{A}, \mathcal{V}, \mathcal{T}^a, \mathcal{T}^r\}$, where $\mathcal{E}$, $\mathcal{R}$ denote the sets of entities and relations, respectively; $\mathcal{A}$ and $\mathcal{V}$ represent the sets of attributes and attribute values. A knowledge graph consists of two categories of triples: attribute triples and relation triples. Attribute triples describe the attributes of entities and are defined as $\mathcal{T}^a=\{(e, a, v)\mid e\in \mathcal{E}, a\in \mathcal{A}, v\in \mathcal{V}\}$. Relation triples describe different relations between entities and are defined as $\mathcal{T}^r=\{(h, r, t)\mid h, t\in \mathcal{E}, r \in \mathcal{R}\}$. 
\end{definition}

\begin{definition}
(Entity Alignment) Given a source knowledge graph $\mathcal{KG}_s=\{\mathcal{E}_s, \mathcal{R}_s, \mathcal{A}_s, \mathcal{V}_s, \mathcal{T}^a_s,\\ \mathcal{T}^r_s\}$ and a target knowledge graph $\mathcal{KG}_t=\{\mathcal{E}_t, \mathcal{R}_t, \mathcal{A}_t, \mathcal{V}_t, \mathcal{T}^a_t, \mathcal{T}^r_t\}$, the task of EA is to identify pairs of entities that refer to the same real-world object across the two graphs. Formally, the goal is to construct the set of aligned entity pairs $\mathcal{S}_{st} = \{(e_i, e_j )\mid e_i\in \mathcal{E}_s, e_j\in \mathcal{E}_t , e_i \Leftrightarrow e_j\}$, where the notation $e_i \Leftrightarrow e_j$ indicates that $e_i$ from $\mathcal{KG}_s$ and $e_j$ from $\mathcal{KG}_t$ share the same semantics and thus correspond to the same real-world object.
\end{definition}

In this work, we consider EA on knowledge graphs as a multi-step decision-making problem. Given a source entity $e_s$ from the source knowledge graph $\mathcal{KG}_s$ and a candidate entity set $\mathcal{C}(e_s)$ retrieved from the target knowledge graph $\mathcal{KG}_t$, the objective is to determine the corresponding target entity $\hat{e}_t \in \mathcal{C}(e_s)$.

\section{Method}

\subsection{Overview}
\begin{figure*}[t]
  \includegraphics[width=\linewidth]{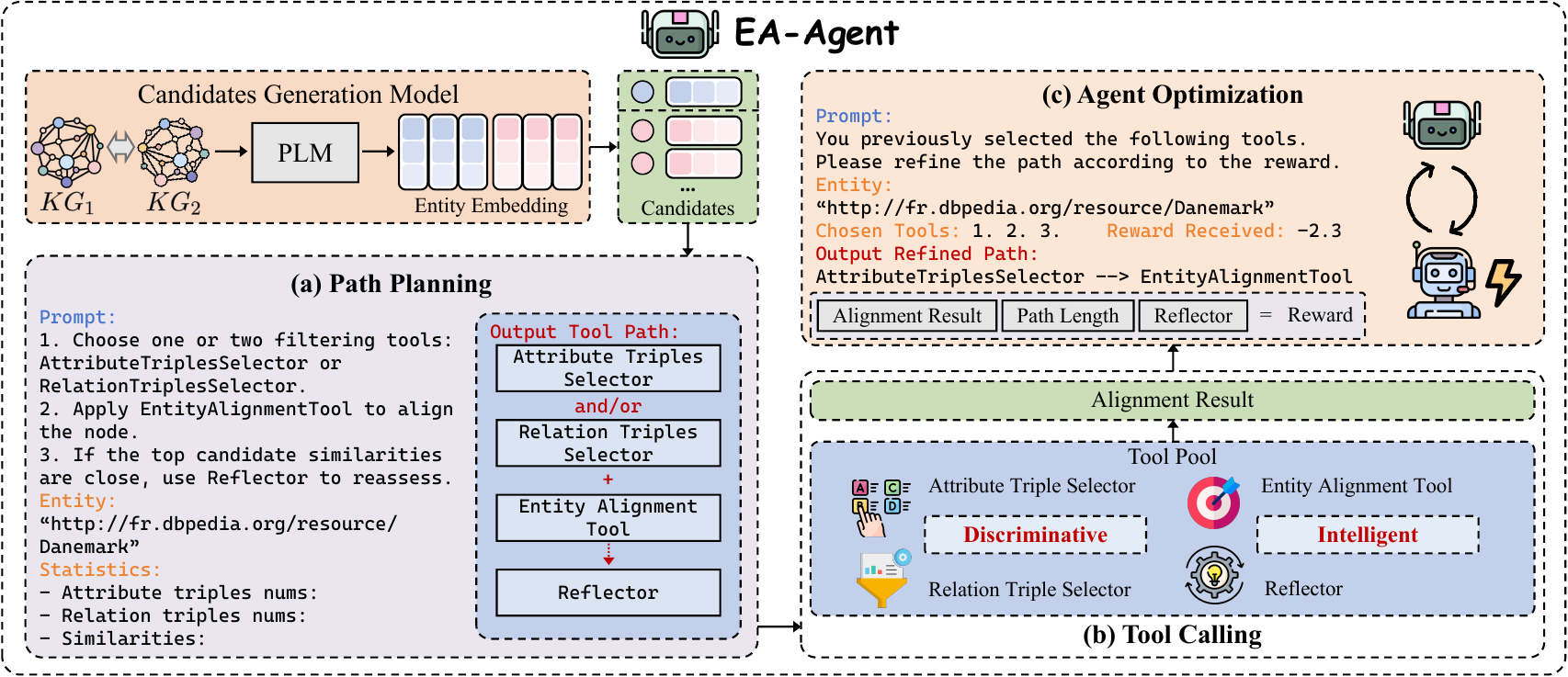}
  \caption {The overview of our proposed EA-Agent, which consists of three main stages: (a) Path Planning stage, (b) Tool Calling stage, and (c) Agent Optimization stage.}
  \label{fig:overview}
\end{figure*}
We propose EA-Agent, a reasoning-driven agent that decomposes EA into a multi-step process of tool planning and execution. The available tools for EA-Agent include an Attribute Triple Selector, a Relation Triple Selector, an Entity Alignment Tool, and a Reflector. Figure \ref{fig:overview} presents the overall architecture of EA-Agent, which consists of three key stages. In the \textbf{Path Planning} stage, EA-Agent observes various statistics of the source entity and the similarity scores of its candidate entities to generate a tool path. In the \textbf{Tool Calling} stage, EA-Agent sequentially invokes these tools specified in the planned path. The Attribute and Relation Triple Selectors distill informative triples to streamline the alignment process, while the Entity Alignment Tool and the Reflector collaborate to generate the initial alignment results. In the \textbf{Agent Optimization} stage, a reward function evaluates both the alignment accuracy and the quality of the tool path. Guided by this reward, EA-Agent is further fine-tuned via the offline policy updating to continuously improve its planning and decision-making capabilities.

\subsection{Path Planning}
To initialize the alignment process, we utilize TEA~\cite{zhao2023alignment} to retrieve the top-$k$ semantically similar entities as candidates, serving as essential inputs for the subsequent reasoning process.

Given the retrieved candidates, EA-Agent performs path planning by jointly considering the structural characteristics of the source entity and the similarity scores of its candidates. The objective of path planning is to enable the agent to autonomously design a multi-step tool path by determining which tools to invoke and in what order. Based on these observed signals, EA-Agent assesses both the informativeness of the entity features and the uncertainty of the candidate set: it first decides whether to invoke attribute or relation triple selectors to filter redundant information, then applies the Entity Alignment Tool, and finally activates the Reflector only when the candidate similarities indicate ambiguity.

To support this decision process, we provide EA-Agent with key statistics extracted from both the source entity and the candidate set:

\begin{EntityStatBox}{Entity Statistics}
    Attribute triples: \texttt{\{attr\_cnt\_all\}} \\
    Attribute types: \texttt{\{attr\_cnt\}} \\
    Relation triples: \texttt{\{rel\_cnt\_all\}} \\
    Relation types: \texttt{\{rel\_cnt\}} \\
    Has name attribute: \texttt{\{signal\_attr\}} \\
    Candidate similarities: top1=\texttt{\{top1\_score\}}, top2=\texttt{\{top2\_score\}}, top3=\texttt{\{top3\_score\}}
\end{EntityStatBox}

This structured reasoning process ensures that each reasoning step plays a clear role in the alignment procedure, enabling EA-Agent to generate an interpretable and adaptive tool path.

\subsection{Tool Calling}
Once the tool path is determined during the Path Planning stage, the system executes each tool step-by-step according to the planned order. EA-Agent provides a tool pool consisting of an Attribute Triple Selector, a Relation Triple Selector, an Entity Alignment Tool, and a Reflector. The execution follows a hierarchical pipeline: the triple selectors first extract the most discriminative structural and semantic information from raw triples, which is then utilized by the Entity Alignment Tool to produce an initial prediction. When the alignment result is uncertain, the Reflector re-evaluates and revises the initial prediction based on prior contextual signals. This Tool Calling stage enables EA-Agent to exert fine-grained control over the information utilized during cross-KG alignment, resulting in more robust and reliable alignment performance. The concrete descriptions of the tools used are given here. 

\begin{itemize}
\item \textbf{Attribute Triple Selector}: It aims to retain the most informative attributes for entity discrimination. It combines an entropy-based criterion with predefined important attributes to avoid losing key semantic signals. The entropy of an attribute $a$ is computed as
\begin{equation}
H(a)=-\sum_{v\in V(a)}p(v)\log p(v),
\end{equation}
where $V(a)$ denotes the set of attribute values and $p(v)$ represents the empirical distribution of value $v$ among the candidate entities. Attributes with lower entropy are considered more discriminative and are preferentially selected, while important attributes are always preserved.

\item \textbf{Relation Triple Selector}: It is inspired by inverse-frequency. Given a relation $r$ with frequency $freq(r)$ in the graph and the total number of relation triples $N$, its score is defined as
\begin{equation}
I(r)=\log\left(\frac{N}{freq(r)+1}\right).
\end{equation}
Rare relations are assigned higher scores since they are more discriminative for EA, similar to the inverse document frequency (IDF) principle in information retrieval.

\item \textbf{Entity Alignment Tool}: It is driven by an LLM, which takes the selected attribute and relation triples as input and outputs the final alignment prediction by jointly reasoning over semantic, attribute, and relational information.

\item \textbf{Reflector}: It is an LLM-based module for result verification and correction. When the alignment is uncertain, it re-evaluates the candidates based on prior context and provides a refined prediction, helping to reduce hallucinations and improve robustness.
\end{itemize}

\subsection{Agent Optimization}
Based on the above two stages, EA-Agent is capable of generating reasonable tool usage paths and completing EA. However, our experiments show that, due to the inherent instability of LLMs, single-round planning can result in redundant or inefficient paths as well as unstable alignment outcomes (See details in Appendix~\ref{app:analysis}.). To address this issue, we introduce an Agent optimization stage that continuously provides feedback and correction signals to EA-Agent through a reward function, enabling the path planning policy to be progressively improved.

\noindent
\textbf{Reward function.} To evaluate the quality of the tool paths and to provide supervision signals for policy optimization, we design a comprehensive reward function. The reward function accounts for not only the correctness of the final alignment but also the efficiency of the tool path and the necessity of Reflector invocation. In this way, EA-Agent is guided to learn a more efficient and stable tool planning strategy while maintaining high alignment accuracy.

Specifically, for each source entity, after the agent completes the tool calling stage, it obtains an alignment prediction $\hat{y}$. The corresponding reward score is composed of three factors:
\begin{equation}
\gamma = \gamma_{\mu}+ c\cdot\gamma_{\text{ref}} + \gamma_{\text{e}}.
\end{equation}

First, $\gamma_{\mu}$ measures the correctness of the final alignment result and serves as the core component of the reward function. If the predicted result $\hat{y}$ matches the ground-truth label $y$, the reward is set to 1; otherwise, it is set to 0:
\begin{equation}
\gamma_{\mu}=\left\{\begin{array}{ll}
1, & \text { if } \hat{y}=y, \\
0, & \text { otherwise. }
\end{array}\right.
\end{equation}

Second, $\gamma_{\text{ref}}$ assesses the rationality of activating the Reflector. This term encourages the Reflector to be activated when necessary to correct erroneous predictions, while suppressing unnecessary or even harmful reflections. Let $\hat{y}$ denote the initial prediction and $\tilde{y}$ denote the refined prediction after reflection. It is defined as:
\begin{equation}
\gamma_{\text{ref}} =
\begin{cases}
1, & \text{if } \hat{y} \neq y\text{ and }\tilde{y} = y, \\
-\alpha, & \text{if } \hat{y} = y\text{ and } \tilde{y} = y, \\
-1, & \text{if } \hat{y} = y\text{ and }\tilde{y} \neq y.
\end{cases}
\end{equation}

When the initial prediction is incorrect and is successfully corrected by the Reflector, a positive reward is assigned; when the initial prediction is already correct but the Reflector is unnecessarily invoked without changing the result, a mild penalty is imposed to discourage redundant reflection; when the Reflector mistakenly alters a correct prediction into an incorrect one, the maximum penalty is applied.

Finally, $\gamma_{\text{e}}$ constrains the computational overhead by penalizing long tool paths. Let the path length be $l$, and we define:
\begin{equation}
\gamma_{\text{e}}=e^{-\beta \cdot l},
\end{equation}
where $\beta$ is a coefficient used to balance alignment accuracy and path complexity. Longer paths lead to larger penalties, thereby encouraging EA-Agent to select simpler and more efficient tool invocation paths while preserving alignment performance.

\noindent
\textbf{Policy Updating.} After obtaining the complete reward signals, we adopt an offline policy updating strategy to continuously optimize the agent’s path planning policy. Specifically, for each tool invocation path, we preserve three types of key information to construct a policy update triple: (1) the initial tool path generated by EA-Agent; (2) the corresponding reward score $\gamma$; and (3) the refined path rewritten by EA-Agent under reward guidance. These policy update triples, which contain preference information, are incorporated into an offline trajectory dataset $\mathcal{D}^*$.
During training, we employ supervised fine-tuning (SFT) to optimize the agent policy model, with the objective function defined as:
\begin{equation}
\mathcal{L}_{\mathrm{SFT}}=-\mathrm{E}_{(q, \mu) \sim \mathcal{D}^{*}}\left[\log \pi_{\theta}(\mu \mid q)\right],
\end{equation}
where $q$ denotes the input context for path planning, $\mu$ denotes the target tool path obtained under reward guidance, and $\pi_{\theta}$ is the agent policy model parameterized by $\theta$.
After each iteration of policy updating, the updated agent is redeployed to execute path planning, tool calling, and reward evaluation, continuously generating new trajectory data for the next iteration of offline optimization. 

\subsection{Complexity Analysis}
EA-Agent is a multi-stage agent whose overall workflow follows a closed-loop iterative process consisting of path planning, tool execution, reward evaluation, and policy updating. At each iteration, the agent plans and executes a tool invocation path, evaluates the resulting trajectory using our proposed reward function, and updates its policy in an offline manner. This forms a closed-loop “planning–execution–evaluation–updating” process that iterates until convergence, enabling EA-Agent to progressively learn more efficient and stable tool planning strategies.
\par
From the perspective of computational complexity, EA-Agent adopts the offline policy training and inference rather than online decision-making, which achieves better training stability and controllable computational cost. In addition, the triple selectors effectively reduce prompt length and inference cost by filtering redundant triples. The complete pseudocode of EA-Agent is provided in Appendix~\ref{app:pseudocode}.

\section{Experiments}
In this section, we evaluate EA-Agent on public datasets. We conduct extensive experiments to demonstrate the effectiveness of our method by answering the following research questions: \textbf{RQ1}: How does EA-Agent perform compared with SOTA baselines on benchmark EA datasets? \textbf{RQ2}: How do different components of EA-Agent affect the final alignment performance? \textbf{RQ3}: Can EA-Agent provide an interpretable alignment process through tool planning and agent optimization? \textbf{RQ4}: How does EA-Agent compare in terms of cost and efficiency to an LLM-based baseline?

\subsection{Experiment Settings}
\subsubsection{Datasets}

We evaluate EA-Agent on the widely used DBP15K dataset~\cite{sun2017cross}, which consists of three cross-lingual EA datasets: FR–EN, JA–EN, and ZH–EN. Additionally, we conduct experiments on the SRPRS dataset~\cite{guo2019learning}, including the EN-FR and EN-DE cross-lingual EA datasets. The statistical characteristics of the datasets are reported in Appendix~\ref{app:dataset_statis}.

\subsubsection{Baseline}
We compare EA-Agent with 10 representative baselines across three categories: translation-based, GNN-based, and PLM-based methods. All baseline models follow the hyperparameter settings reported in their original papers. For brevity, we detail the specific models in Appendix~\ref{app:baseline}.

\subsubsection{Implement Details}
We use TEA~\cite{zhao2023alignment} as the SLM to generate Top-10 candidate entity lists, due to its strong Hits@10 performance. We adopt Qwen3-32B, Qwen3-8B~\cite{yang2025qwen3}, and Llama3-8B-Instruct~\cite{dubey2024llama} as backbone LLMs for EA-Agent.
The agent is optimized using LoRA with the AdamW optimizer for three epochs. For fair comparison with prior work, all datasets are split into training and test sets with a ratio of 3:7. All experiments are conducted on two NVIDIA A100 80G GPUs.
Detailed hyperparameters are provided in Appendix~\ref{sec:appendix1}.

\subsection{Performance Comparison (RQ1)}
\begin{table*}[t]
\footnotesize
\centering
\caption{Overall entity alignment performance on the DBP15K datasets.}
\label{tab:main_results}
\renewcommand{\arraystretch}{1.1}
\begin{tabular}{l|ccc|ccc|ccc}
\hline
\multirow{2}{*}{Method} 
& \multicolumn{3}{c|}{FR-EN} 
& \multicolumn{3}{c|}{JA-EN} 
& \multicolumn{3}{c}{ZH-EN}  \\
& Hits@1 & MRR & Hits@10 
& Hits@1 & MRR & Hits@10 
& Hits@1 & MRR & Hits@10 \\
\hline
MTransE     & 24.40 & 0.335 & 55.6 & 27.90 & 0.349 & 57.5 & 30.80 & 0.364 & 61.4 \\
JAPE        & 32.40 & 0.430 & 66.7 & 36.30 & 0.476 & 68.5 & 41.20 & 0.490 & 74.5 \\
GCN-Align   & 37.30 & 0.532 & 74.5 & 39.90 & 0.546 & 74.5 & 41.30 & 0.549 & 74.4 \\
RDGCN       & 88.60 & 0.911 & 95.7 & 76.70 & 0.812 & 89.5 & 70.80 & 0.746 & 84.6 \\
BERT-INT    & 98.70 & 0.990 & 99.2 & 80.60 & 0.820 & 83.5 & 81.40 & 0.820 & 83.7 \\
SDEA        & 96.60 & 0.980 & 99.5 & 84.80 & 0.890 & 95.2 & 87.00 & 0.910 & 96.6 \\
TEA         & 95.80 & 0.970 & 99.7 & 89.00 & 0.920 & 97.8 & 84.50 & 0.887 & 98.2 \\
LLMEA       & 95.70 & 0.957 & --   & 91.10 & 0.911 & --   & 89.80 & 0.898 & --   \\
ChatEA      & 99.00 & 0.995 & --   & --    & --    & --   & --    & --    & -- \\
Seg-Align   & 98.70 & 0.987 & --   & 90.70 & 0.907 & --   & 95.30 & 0.953 & --    \\
\hline
\textbf{EA-Agent} 
& \textbf{99.00} & \textbf{0.996} & \textbf{99.7}$^{*}$ 
& \textbf{94.27} & \textbf{0.950} & \textbf{97.8}$^{*}$ 
& \textbf{96.67} & \textbf{0.971} & \textbf{98.2}$^{*}$ \\
\hline
\end{tabular}
\footnotesize

$^{*}$Hits@10 of EA-Agent is determined by the Top-10 candidates retrieved by TEA.
\end{table*}

\begin{table}[t]
\footnotesize
\centering
\caption{Overall entity alignment performance on the SRPRS datasets.}
\label{tab:srprs_results}
\renewcommand{\arraystretch}{1.1}
\begin{tabular}{l|cc|cc}
\hline
\multirow{2}{*}{Method} 
& \multicolumn{2}{c|}{EN-FR} 
& \multicolumn{2}{c}{EN-DE} \\
& Hits@1 & MRR 
& Hits@1 & MRR \\
\hline
MTransE     & 21.3 & 0.29 & 10.7 & 0.16 \\
JAPE        & 24.1 & 0.34 & 26.8 & 0.36 \\
GCN-Align   & 29.6 & 0.40 & 42.8 & 0.51 \\
RDGCN       & 67.2 & 0.71 & 77.9 & 0.82 \\
BERT-INT    & 97.1 & 0.97 & 98.6 & 0.99 \\
SDEA        & 96.6 & 0.97 & 96.8 & 0.98 \\
TEA         & 98.5 & 0.99 & 98.7 & 0.99 \\
Seg-Align   & 98.2 & 0.982 & 98.8 & 0.988 \\
\hline
\textbf{EA-Agent} 
& \textbf{99.0} & \textbf{0.996} 
& \textbf{99.5} & \textbf{0.997} \\
\hline
\end{tabular}
\end{table}

Table \ref{tab:main_results} reports the overall performance of EA-Agent on the DBP15K datasets. In all experiments, we use TEA to generate Top-10 candidate entities and adopt Qwen3-32B as the backbone agent. Thus, the reported Hits@10 represents the recall performance of the PLM-based EA model.

The results for RQ1 show that EA-Agent consistently outperforms state-of-the-art EA methods across all datasets. EA-Agent achieves competitive performance on FR–EN and establishes new SOTA results on JA–EN and ZH–EN, exceeding the strongest baselines by 3.17 and 1.37 Hits@1 points, respectively. Despite TEA already providing high-recall candidate sets, the consistent gains in Hits@1 and MRR indicate that EA-Agent is effective at selecting the correct entity from high-quality candidates.

We compare EA-Agent with LLM-based baselines: LLMEA (ERNIE), ChatEA (GPT-4), and Seg-Align (GPT-3.5). Their strong performance largely derives from the underlying LLMs' inherent reasoning capacity. Conversely, EA-Agent achieves comparable performance using open-source models, demonstrating that gains primarily stem from our proposed tool planning and optimization rather than the LLM itself.

To further evaluate the robustness of EA-Agent, we conduct experiments on the SRPRS dataset, which is widely used to assess entity alignment performance under more challenging settings. The results are shown in Table \ref{tab:srprs_results}.

\subsection{Ablation Study (RQ2)}
To answer RQ2 and systematically evaluate the contribution of each component in EA-Agent, we conduct a series of ablation studies on the DBP15K benchmark. The ablation experiments focus on three key aspects of the agent: \textbf{the Path Planning stage and the associated triple selectors}, \textbf{different LLMs} used as the agent backbone, and \textbf{the Agent Optimization stage}. 

\begin{table}[t]
\centering
\caption{Ablation study on the impact of different components in EA-Agent.}
\footnotesize
\setlength{\tabcolsep}{5pt}
\renewcommand{\arraystretch}{1.1}
\begin{tabular}{lcc|cc}
\toprule
\multirow{2}{*}{\textbf{Method}} 
& \multicolumn{2}{c|}{\textbf{FR--EN}} 
& \multicolumn{2}{c}{\textbf{JA--EN}} \\
& Hits@1 & MRR & Hits@1 & MRR \\
\midrule
All Components 
& \textbf{97.30} & \textbf{0.975} 
& \textbf{90.67} & \textbf{0.913} \\
\midrule
Rule-based 
& 95.65 & 0.962  
& 88.29 & 0.890 \\
\textit{w/o} Path Planning 
& 94.85 & 0.950 
& 87.29 & 0.880 \\
\textit{w/o} Selector      
& 93.34 & 0.935 
& 86.75 & 0.847 \\
LLM-Only                  
& 89.44 & 0.901
& 83.50 & 0.841 \\
\bottomrule
\end{tabular}
\label{tab:ablation}
\end{table}

\noindent
\textbf{Path Planning stage.} We evaluate the path planning and triple selection mechanisms on the FR–EN and JA–EN datasets using three ablated variants: (1) LLM-Only, which performs EA without triples; (2) w/o Selector, which feeds all triples without filtering; (3) w/o Path Planning, which invokes all tools without adaptive planning; and (4) Rule-based, a variant that triggers the Reflector only when the similarity gap between the top-2 candidates is below a threshold (0.3). The first two variants assess the impact of triple selection, while the latter two evaluate the role of adaptive path planning.

As shown in Table \ref{tab:ablation}, directly using an LLM for EA performs poorly, indicating that structured triples provide essential alignment signals. Although using all triples improves performance, it also introduces noise and token redundancy. Moreover, the Rule-based variant achieves 95.65\% Hits@1 on FR-EN, which is lower than the full EA-Agent. This result, along with the performance degradation of the w/o Path Planning variant, indicates that our planning module provides additional benefits beyond static threshold-based strategies and prevents the overuse of the Reflector, which could otherwise lead to erroneous corrections.

\begin{table}[t]
\centering
\caption{Performance (\%) of EA-Agent with different LLM backbones on DBP15K.}
\label{tab:llm_backbone}
\footnotesize
\setlength{\tabcolsep}{3pt}
\renewcommand{\arraystretch}{1.1}
\begin{tabular}{l|c c c}
\toprule
\textbf{LLM} 
& \textbf{FR--EN} 
& \textbf{JA--EN} 
& \textbf{ZH--EN} \\
\midrule
Qwen3-8B 
& 88.10 / .887
& 85.95 / .867 
& 83.63 / .840 \\
Llama3-8B-Instruct 
& 92.34 / .929
& 89.69 / .901 
& 90.13 / .907 \\
Qwen3-32B 
& \textbf{99.00} / \textbf{.996} 
& \textbf{94.27} / \textbf{.950} 
& \textbf{96.67} / \textbf{.971} \\
\bottomrule
\end{tabular}
\end{table}

\noindent
\textbf{Different LLMs as backbone.} We further evaluate EA-Agent with three backbone LLMs—Qwen3-32B, Qwen3-8B, and Llama-3-8B-Instruct—on the FR–EN, JA–EN, and ZH–EN datasets. The experimental results are reported in Table \ref{tab:llm_backbone}. EA-Agent achieves the best performance with Qwen3-32B, and the comparison with Qwen3-8B indicates a clear scaling-law effect. Llama-3-8B-Instruct slightly outperforms Qwen3-8B, likely due to its instruction tuning that enhances agent reasoning.

\begin{figure}[t]
    \centering
    \includegraphics[width=\columnwidth]{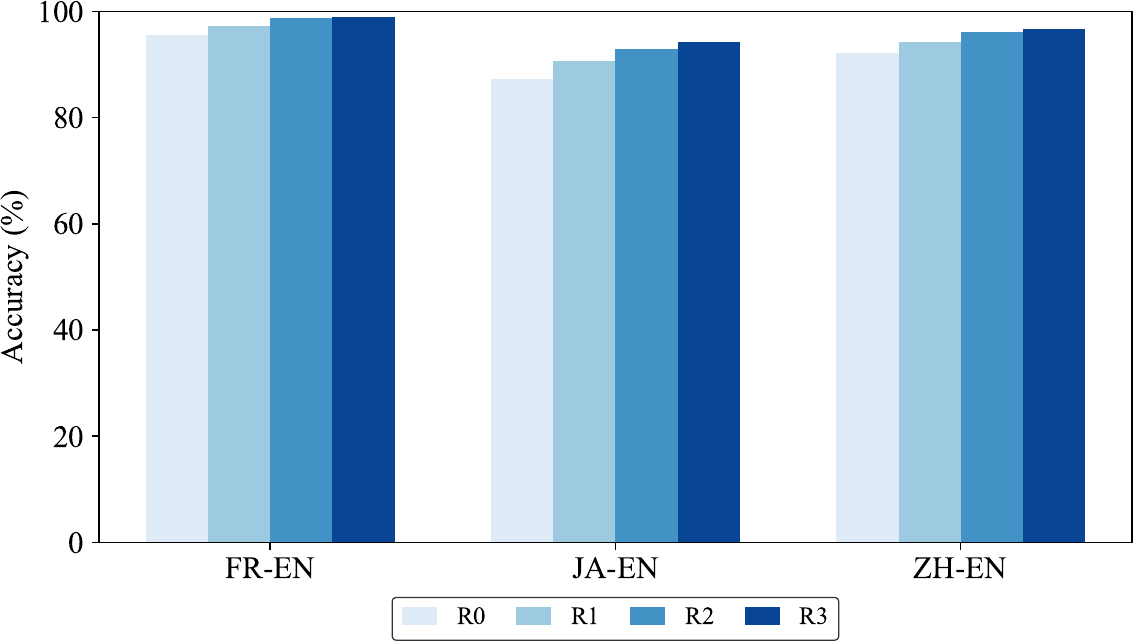}
    \caption{Accuracy improvements over multiple optimization rounds on the FR--EN, JA--EN, and ZH--EN datasets. Each group of bars corresponds to a dataset, where colors from light to dark denote Round 0 to Round 3, respectively.}
    \label{fig:round_accuracy}
\end{figure}

\noindent
\textbf{Agent Optimization stage.} We analyze EA-Agent across multiple optimization rounds on the FR–EN, JA–EN, and ZH–EN datasets, as shown in Figure~\ref{fig:round_accuracy}. Alignment accuracy consistently improves with successive rounds, demonstrating that the reward function effectively guides EA-Agent toward better tool-planning strategies. The gains gradually diminish after the second round, indicating convergence to a stable and near-optimal policy. Based on this observation, we adopt three optimization rounds in all experiments to balance performance and computational cost.

\subsection{Case Study (RQ3)}
To further illustrate the interpretability and the effectiveness of both the tool planning and Agent Optimization stages, we present a representative case study from the FR-EN dataset.

Given a source entity "\url{http://fr.dbpedia.org/resource/Saint-Isidore\_(Roussillon)}" in the source knowledge graph, EA-Agent first generates a tool path at Round 0 as \textit{AttributeTripleSelector → RelationTripleSelector → EntityAlignmentTool → Reflector}. Following this, the Entity Alignment Tool produces an initial prediction "\url{http://dbpedia.org/resource/Saint-Isidore,\_Montérégie,\_Quebec}". However, the Reflector incorrectly revises it to an incorrect final alignment result "\url{http://dbpedia.org/resource/Saint-Isidore,\_Chaudière-Appalaches,\_Quebec}". Due to unnecessary reflector invocation, EA-Agent receives a low reward. During the agent optimization phase, guided by the reward signal, the agent rewrites the path as \textit{AttributeTripleSelector → RelationTripleSelector → EntityAlignmentTool}. In the subsequent round, the optimized path produces the correct target entity and achieves a higher reward.

This case demonstrates that explicit tool planning yields transparent reasoning paths, while reward-guided optimization effectively eliminates suboptimal tool usage, improving both alignment accuracy and efficiency.

\subsection{Efficiency Analysis (RQ4)}
We further analyze the efficiency of EA-Agent from three aspects: token cost, inference time, and training cost.

\begin{table}[t]
\centering
\caption{Average token consumption per entity for different LLM-based EA methods.}
\label{tab:token_cost}
\footnotesize
\setlength{\tabcolsep}{8pt}
\begin{tabular}{l|c}
\toprule
Method & Avg. Tokens \\
       & per Entity  \\
\midrule
ChatEA    & 9,803 \\
Seg-Align & 159   \\
EA-Agent    & 672 \\
\bottomrule
\end{tabular}
\end{table}

Table \ref{tab:token_cost} reports the average token consumption per entity for different LLM-based EA methods. Seg-Align reduces token usage by discarding triples and using only entity names, sacrificing accuracy. In contrast, EA-Agent leverages informative attribute and relation triples while consuming only 672 tokens, significantly less than ChatEA, which also considers triples. This efficiency gain mainly benefits from the Path Planning stage and the triple selectors that filter redundant triples before LLM invocation, allowing EA-Agent to better balance efficiency and effectiveness.

\begin{table}[t]
\centering
\caption{Average time cost per entity (path planning + entity alignment) on different datasets.}
\label{tab:time_cost}
\footnotesize
\setlength{\tabcolsep}{4pt}
\begin{tabular}{c|c c c}
\toprule
Dataset & FR-EN & JA-EN & ZH-EN \\
\midrule
Time (s) & 1.53+2.35 & 1.50+2.48 & 1.55+2.68 \\
\bottomrule
\end{tabular}
\end{table}

We further analyzed the average time cost per entity on the FR-EN, JA-EN, and ZH-EN datasets, as shown in Table \ref{tab:time_cost}. These results indicate that EA-Agent is highly efficient in practice, with a short end-to-end processing time per entity, making it suitable for large-scale EA scenarios.

\begin{table}[t]
\centering
\caption{The comparison of average inference time cost per entity on FR-EN dataset.}
\label{tab:com_cost}
\footnotesize
\setlength{\tabcolsep}{8pt}
\begin{tabular}{l|c}
\toprule
Method & Avg. Infer Time \\
       & per Entity (s)  \\
\midrule
ChatEA    & 4.35 \\
Seg-Align & 0.89   \\
EA-Agent    & 3.88 \\
\bottomrule
\end{tabular}
\end{table}

The comparison of average inference time per entity on the FR-EN dataset are shown in Table \ref{tab:com_cost}. The experimental results show that EA-Agent has an acceptable inference time. Although Seg-Align has a shorter inference time, its accuracy is lower because it does not use triples.

Regarding training efficiency, EA-Agent employs parameter-efficient fine-tuning with LoRA and updates only a small subset of backbone parameters. Agent optimization is performed offline on collected trajectories, avoiding costly online reinforcement learning. Consequently, training remains affordable even with large backbone models and typically converges within a few iterations.

\section{Conclusion}
In this paper, we propose EA-Agent, a reasoning-driven agent that addresses EA through structured multi-step planning and tool execution, overcoming the limitations of existing LLM-based methods in interpretability and efficiency. By integrating explicit path planning, selective tool calling, and reward-guided agent optimization, EA-Agent enables transparent and effective alignment decisions. Extensive experiments on three public datasets demonstrate that EA-Agent consistently achieves state-of-the-art performance.

\section*{Limitations}

Despite its effectiveness, EA-Agent has several limitations. First, the set of tools used by the agent is manually designed based on prior knowledge of the entity alignment task. While these tools are sufficient to achieve strong performance in our experiments, extending EA-Agent to automatically design tool sets remains an interesting direction for future work. Second, the agent optimization in EA-Agent is performed in an offline and supervised manner using reward-guided trajectory rewriting, rather than end-to-end reinforcement learning. While this design improves training stability, it may limit exploration in more complex long-horizon planning scenarios.

\section*{Acknowledgments}
The authors would like to thank the Area Chair and the anonymous reviewers for their insightful comments and constructive suggestions, which helped to improve the quality of this paper.


\bibliography{main}

\appendix

\section{Analysis of Single-Round Planning}
\label{app:analysis}

\begin{figure}[bp]
  \centering
  \includegraphics[width=\columnwidth]{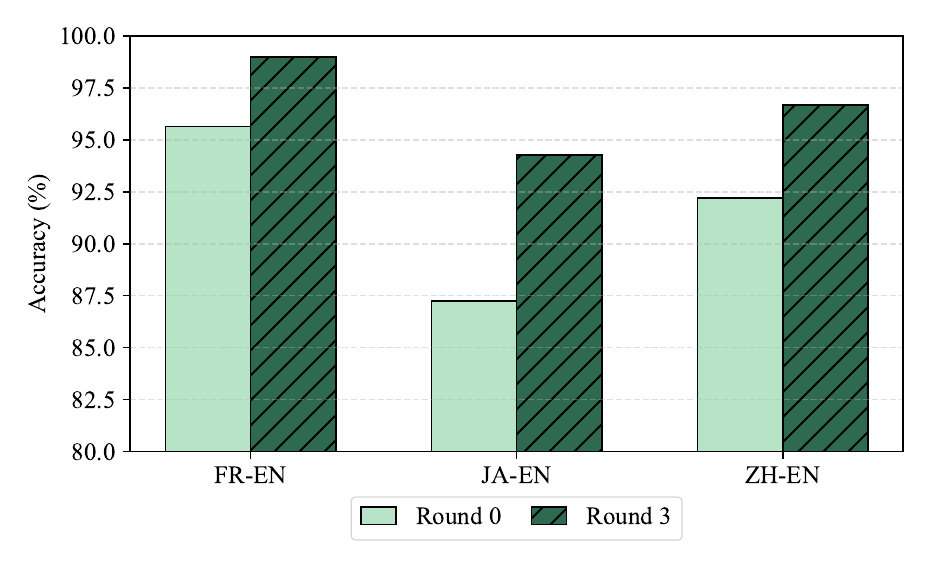}
  \caption{Alignment accuracy across different planning rounds.}
  \label{fig:round_acc}
\end{figure}

To further analyze the stability and efficiency of the planning strategy, we compare single-round planning (Round 0) with optimized planning after multiple optimization rounds (Round 3). Specifically, we evaluate the Recall of the Reflector and the average tool path length under the two settings, as reported in Table~\ref{tab:reflector_recall} and Table~\ref{tab:path_length}, respectively. In addition, the alignment accuracy across planning rounds is illustrated in Figure~\ref{fig:round_acc}.

The results reveal clear limitations of single-round planning. As shown in Table~\ref{tab:reflector_recall}, the Reflector recall in Round~0 is significantly higher than that in Round~3, indicating that EA-Agent frequently resorts to reflection due to uncertainty in early-stage planning. Meanwhile, Table~\ref{tab:path_length} shows that single-round planning produces longer tool paths on average, suggesting the presence of redundant and inefficient tool invocations. Consistently, Figure~\ref{fig:round_acc} demonstrates that alignment accuracy under single-round planning is notably lower than that achieved after optimization. These observations indicate that planning based solely on LLM is prone to redundancy and instability, leading to suboptimal alignment results.

\begin{table}[bp]
\centering
\caption{Comparison of the Reflector recall between single-round planning (Round 0) and optimized planning (Round 3).}
\footnotesize
\setlength{\tabcolsep}{6pt}
\begin{tabular}{lcc}
\toprule
\textbf{Planning Round} & \textbf{Reflector Recall (\%)} \\
\midrule
Round 0 & 90.07 \\
Round 3 & 19.69 \\
\bottomrule
\end{tabular}
\label{tab:reflector_recall}
\end{table}

\begin{table}[bp]
\centering
\footnotesize
\caption{Average tool path length under different planning rounds.}
\setlength{\tabcolsep}{6pt}
\begin{tabular}{lcc}
\toprule
\textbf{Planning Round} & \textbf{Avg. Path Length} \\
\midrule
Round 0 & 3.61 \\
Round 3 & 2.98 \\
\bottomrule
\end{tabular}
\label{tab:path_length}
\end{table}

\section{Pseudocode of the EA-Agent Framework}
\label{app:pseudocode}
In this section, we illustrate the pseudocode of EA-Agent during the training and testing phases in Algorithm \ref{alg:EA-Agent-train} and Algorithm \ref{alg:EA-Agent-test}, respectively.

\begin{algorithm}[t]
\footnotesize
\caption{EA-Agent Training Phase}
\label{alg:EA-Agent-train}
\KwIn{
Training entities $\mathcal{E}_{\text{train}}$;
candidate set $\mathcal{C} \subset \mathcal{KG}_t$; 
initial policy $\pi_0$
}
\KwOut{Optimized agent policy $\pi$}

Initialize policy $\pi \leftarrow \pi_0$\;

\For{round $r = 1,\dots,R$}{
    Initialize trajectory dataset $\mathcal{D}_r \leftarrow \emptyset$\;
    
    \For{each $e_s \in \mathcal{E}_{\text{train}}$}{
        Retrieve candidate set $\mathcal{C}(e_s) \subset \mathcal{C}$\;
        Compute statistics $\mathcal{S}(e_s)$\;

        Construct query $q = (\mathcal{S}(e_s), \mathcal{C}(e_s))$\;
        Sample tool path $\mu \sim \pi(\cdot \mid q)$\;

        Execute $\mu$ to obtain prediction $\hat{e}_t$\;
        Compute reward $\gamma$ according to Eq.~(3)\;

        $\mathcal{D}_r \leftarrow \mathcal{D}_r \cup \{(q, \mu, \gamma)\}$\;
    }

    Update policy $\pi$ by minimizing $\mathcal{L}_{\mathrm{SFT}}$ in Eq.~(7)\;
}

\Return{$\pi$}
\end{algorithm}

\begin{algorithm}[t]
\footnotesize
\caption{EA-Agent Inference Phase}
\label{alg:EA-Agent-test}
\KwIn{
Test entities $\mathcal{E}_{\text{test}}$;
candidate set $\mathcal{C} \subset \mathcal{KG}_t$; 
trained policy $\pi$
}
\KwOut{Aligned entities $\hat{e}_t$}

\For{each $e_s \in \mathcal{E}_{\text{test}}$}{
    Retrieve candidate set $\mathcal{C}(e_s) \subset \mathcal{C}$\;
    Compute statistics $\mathcal{S}(e_s)$\;

    Construct query $q = (\mathcal{S}(e_s), \mathcal{C}(e_s))$\;
    Sample tool path $\mu \sim \pi(\cdot \mid q)$\;

    Execute $\mu$ to obtain aligned entity $\hat{e}_t$\;
}

\Return{$\hat{e}_t$}
\end{algorithm}

\section{Baseline Details}
\label{app:baseline}
We compare EA-Agent with 10 representative EA methods covering different categories. The translation-based methods include MTransE~\cite{chen2016multilingual} and JAPE~\cite{sun2017cross}, the GNN-based methods include GCN-Align~\cite{wang2018cross} and RDGCN~\cite{wu2019relation}, and the PLM–based methods include BERT-INT~\cite{tang2020bert}, SDEA~\cite{zhong2022semantics}, TEA~\cite{zhao2023alignment}, LLMEA~\cite{yang2024two}, ChatEA~\cite{jiang2024unlocking}, and Seg-Align~\cite{yang2024advancing}, among which LLMEA, ChatEA, and Seg-Align are LLM-based EA approaches. All baseline models follow the hyperparameter settings reported in their original papers.

\section{Dataset Statistics}
\label{app:dataset_statis}
DBP15K is constructed from DBpedia, a large-scale multilingual knowledge graph that contains abundant inter-lingual links across different language editions. Three cross-lingual EA benchmark datasets are derived from its subgraphs, namely French–English (FR–EN), Japanese–English (JA–EN), and Chinese–English (ZH–EN).

Table~\ref{tab:dbp15k_stats} reports the detailed statistics of the DBP15K datasets used in our experiments.

\begin{table}[t]
\centering
\caption{Statistics of the DBP datasets.}
\label{tab:dbp_stats}
\scriptsize
\setlength{\tabcolsep}{4pt}
\resizebox{\linewidth}{!}{
\begin{tabular}{lcccccc}
\toprule
Dataset & Lang. & Entity & Rel. & Attr. & Rel. triples & Attr. triples \\
\midrule
\multirow{2}{*}{FR--EN} 
 & (FR) & 19,661 & 903 & 4,431 & 105,998 & 528,665 \\
 & (EN) & 19,993 & 1,208 & 6,161 & 115,722 & 576,543 \\
\midrule
\multirow{2}{*}{JA--EN} 
 & (JA) & 19,814 & 1,299 & 5,681 & 77,214 & 354,619 \\
 & (EN) & 19,780 & 1,153 & 5,850 & 93,484 & 497,230 \\
\midrule
\multirow{2}{*}{ZH--EN} 
 & (ZH) & 19,388 & 1,701 & 7,780 & 70,414 & 379,684 \\
 & (EN) & 19,572 & 1,323 & 6,933 & 95,142 & 567,755 \\
\bottomrule
\end{tabular}
}
\end{table}

\begin{table}[t]
\centering
\caption{Statistics of the SRPRS datasets.}
\label{tab:srprs_stats}
\scriptsize
\setlength{\tabcolsep}{4pt}
\resizebox{\linewidth}{!}{
\begin{tabular}{lcccccc}
\toprule
Dataset & Lang. & Entity & Rel. & Attr. & Rel. triples & Attr. triples \\
\midrule
\multirow{2}{*}{EN--FR} 
 & (EN) & 15,000 & 221 & 274 & 36,508 & 70,750 \\
 & (FR) & 15,000 & 177 & 393 & 33,532 & 56,344 \\
\midrule
\multirow{2}{*}{EN--DE} 
 & (EN) & 15,000 & 222 & 275 & 38,363 & 62,715 \\
 & (DE) & 15,000 & 120 & 185 & 37,377 & 142,506 \\
\bottomrule
\end{tabular}
}
\end{table}

\section{Detailed Implementation Hyperparameters}
\label{sec:appendix1}
For LoRA-based fine-tuning, we set the rank to $r=4$ with a scaling factor $\alpha=8$. LoRA is applied only to the query and value projection matrices ($q_{proj}$, $v_{proj}$) in the attention layers. The dropout rate is set to 0.05, and no bias parameters are updated. We train EA-Agent for three epochs using the AdamW optimizer with an initial learning rate of $2\times10^{-4}$.

All input sequences are tokenized using the official tokenizer of each backbone model, with truncation and padding to a maximum length of 1024 tokens. The per-device training batch size is set to 1, and gradient accumulation with 8 steps is adopted to form a larger effective batch size. Mixed-precision training with bfloat16 is enabled.

In the Tool Calling stage, both the attribute and relation triple selectors retain at most 5 triples. The path efficiency coefficient in the reward function is set to $\beta=0.2$, and the mild penalty coefficient for the Reflector is set to $\alpha=0.5$. These hyper-parameter values are determined through sensitivity analysis. During inference, deterministic decoding is used for Llama3-8B-Instruct, while a temperature of 0.1 is applied to Qwen models to reduce decoding variance. All experimental results are reported as the average of multiple independent runs. 

\section{Prompt Templates for EA-Agent}
To concretely instantiate the proposed EA-Agent framework, we design a set of structured prompts that govern path planning, tool calling, reflection, and policy optimization. For clarity and reproducibility, we present all prompt templates used in EA-Agent in this appendix.

\begin{PromptBox}{colorgray}{Tool Pool Definition}
    Tool\_Pool = [ \\
    \quad \{ \\
    \quad\quad \textbf{"name":} "\textit{AttributeTripleSelector}", \\
    \quad\quad \textbf{"definition":} "Selects important attribute triples of an entity by removing common or uninformative attributes.", \\
    \quad\quad \textbf{"usage":} "\textit{AttributeTripleSelector}[entity] → list of (entity, attribute, value)" \\
    \quad \}, \\
    \quad \{ \\
    \quad\quad \textbf{"name":} "\textit{RelationTripleSelector}", \\
    \quad\quad \textbf{"definition":} "Selects informative relation triples (outgoing and incoming) based on their distinctiveness.", \\
    \quad\quad \textbf{"usage":} "\textit{RelationTripleSelector}[entity] → list of (subject, relation, object)" \\
    \quad \}, \\
    \quad \{ \\
    \quad\quad \textbf{"name":} "\textit{EntityAlignmentTool}", \\
    \quad\quad \textbf{"definition":} "Aligns a source entity with the most similar target entity from a candidate list.", \\
    \quad\quad \textbf{"usage":} "\textit{EntityAlignmentTool}[source\_\\entity, candidates] → best target entity" \\
    \quad \}, \\
    \quad \{ \\
    \quad\quad \textbf{"name":} "\textit{Reflector}", \\
    \quad\quad \textbf{"definition":} "Reevaluates the alignment result and suggests a better match if needed.", \\
    \quad\quad \textbf{"usage":} "\textit{Reflector}[source\_entity, candidates, initial\_alignment] → confirmed or revised target" \\
    \quad \}]
\end{PromptBox}

\begin{PromptBox}{color1}{Source Entity Prompt}
    \textbf{Descriptions:} This prompt instructs the agent to perform path planning for entity alignment by selecting an appropriate sequence of tools.
    \medskip
    \hrule
    \medskip
    You are an expert in \textbf{Entity Alignment}. \\
    \\
    \textbf{Steps:} \\
    1. Choose one or two filtering tools: \textit{AttributeTripleSelector}, \textit{RelationTripleSelector}. \\
    2. Apply \textit{EntityAlignmentTool} to align the entity. \\
    3. If the top candidate similarities are close, use \textit{Reflector} to reassess. \\
    \\
    \textbf{Available tools:} \\
    \texttt{\{tool\_pool\}} \\
    \\
    \textbf{Entity:} \texttt{\{entity\_iri\}} \\
    \textbf{Statistics:} \\
    - Attribute triples: \texttt{\{attr\_cnt\_all\}} \\
    - Attribute types: \texttt{\{attr\_cnt\}} \\
    - Relation triples: \texttt{\{rel\_cnt\_all\}} \\
    - Relation types: \texttt{\{rel\_cnt\}} \\
    - Has name attribute: \texttt{\{signal\_attr\}} \\
    - Top-1 sim: \texttt{\{top1\_score\}}, Top-2 sim: \texttt{\{top2\_score\}}, Top-3 sim: \texttt{\{top3\_score\}} \\
    \\
    Output the sequence numbers and tool names only, one per line: \\
    1. <ToolName> \\
    2. <ToolName> \\
    3. <ToolName> (optional) \\
    4. <ToolName> (optional)
\end{PromptBox}

\begin{PromptBox}{color2}{Entity Alignment Prompt}
    \textbf{Descriptions:} This prompt instructs the LLM-based Entity Alignment Tool to identify the corresponding target entity from a candidate set.
    \medskip
    \hrule
    \medskip
    You are given a source entity and several candidate entities from another knowledge graph. Each entity is represented as triples (subject, predicate, object). Candidates are sorted by similarity. \\
    \\
    \textbf{Entity:} \texttt{\{source\_iri\}} \\
    \textbf{Triples:} \\
    \texttt{\{source\_triples\}} \\
    \\
    \textbf{Candidates:} \\
    \texttt{\{candidate\_blocks\}} \\
    \\
    Please enter the IRI that best matches the candidate entity using the following format: \\
    \texttt{[IRI]}
\end{PromptBox}

\begin{PromptBox}{color3}{Reflection Prompt}
    \textbf{Descriptions:} This prompt enables the Reflector to verify and reassess the initial alignment result.
    \medskip
    \hrule
    \medskip
    You are performing an entity alignment task. Given a source entity and several candidate target entities (with their triples), you previously selected one of them. \\
    \\
    Now, reflect on whether that choice was optimal. \\
    \\
    \textbf{Entity:} \texttt{\{source\_iri\}} \\
    \textbf{Triples:} \\
    \texttt{\{source\_triples\}} \\
    \\
    \textbf{Candidates:} \\
    \texttt{\{candidate\_blocks\}} \\
    \\
    \textbf{Initial choice:} \texttt{\{initial\_choice\}} \\
    \\
    Is this the best match? \\
    - If yes, return it. \\
    - If not, return a better one. \\
    \\
    Please enter the best matching IRI using the following format: \\
    \texttt{[IRI]}
\end{PromptBox}

\begin{PromptBox}{color4}{Path Rewriting Prompt}
    \textbf{Descriptions:} This prompt instructs the agent to rewrite the previous tool path based on observed reward feedback.
    \medskip
    \hrule
    \medskip
    You are optimizing the tool selection process for an entity alignment task. \\
    \\
    \textbf{Entity:} \texttt{\{entity\}} \\
    \textbf{Candidate Similarities:} \texttt{\{top1\_score\}}, \texttt{\{top2\_score\}}, \texttt{\{top3\_score\}} \\
    \textbf{Previous Tools:} \texttt{\{old\_tools\}} \\
    \textbf{Reward:} \texttt{\{reward\}} \\
    \\
    \textbf{Available tools:} \\
    - \textit{AttributeTripleSelector} \\
    - \textit{RelationTripleSelector} \\
    - \textit{EntityAlignmentTool} \\
    - \textit{Reflector} (use only when similarities are close) \\
    \\
    Please generate an improved sequence of tools. \\
    \\
    Output the sequence numbers and tool names only, one per line:\\
    1. <ToolName> \\
    2. <ToolName> \\
    3. <ToolName> (optional) \\
    4. <ToolName> (optional)
\end{PromptBox}

\end{document}